\documentclass[twocolumn,prl,amsmath,amssymb]{revtex4-1}
\usepackage{bm}
\newcommand{\tr}{\operatorname{tr}}
\begin{document}

\title{Symmetric Informationally-Complete Quantum States as Analogues to Orthonormal
Bases and Minimum-Uncertainty States}

\author{D.~M. Appleby$^1$, Hoan Bui Dang$^{1,2}$, and Christopher A. Fuchs$^1$}

\affiliation{$^1$Perimeter Institute for Theoretical Physics, Waterloo, Ontario N2L 2Y5, Canada \\
$^2$Institute for Quantum Computing, University of Waterloo, Waterloo, Ontario N2L
3G1, Canada}

\begin{abstract}
Since Renes et al.\ [J. Math.\ Phys.\ {\bf 45}, 2171 (2004)], there
has been much effort in the quantum information community to prove
(or disprove) the existence of symmetric informationally complete
(SIC) sets of quantum states in arbitrary finite dimension. This
paper strengthens the urgency of this question by showing that {\it
if\/} SIC-sets exist: 1) by a natural measure of orthonormality, they
are as close to being an orthonormal basis for the space of density
operators as possible, and 2) in prime dimensions, the standard
construction for complete sets of mutually unbiased bases and
Weyl-Heisenberg covariant SIC-sets are intimately related: The latter
represent minimum uncertainty states for the former in the sense of
Wootters and Sussman. Finally, we contribute to the question of
existence by conjecturing a quadratic redundancy in the equations for
Weyl-Heisenberg SIC-sets.
\end{abstract}

\maketitle

\section{Introduction}
Recently there has been significant interest in the
quantum-information community to prove or disprove the general
existence of so-called {\it symmetric informationally complete\/}
(SIC) quantum measurements \cite{Zauner,Caves,Renes,JMP,SICGeneral,ScottA,Minsk}.
The question is simple enough: For a Hilbert space ${\cal H}_d$ of
arbitrary dimension $d$, is there always a set of $d^2$ quantum
states $|\psi_i\rangle$ such that
$|\langle\psi_i|\psi_j\rangle|^2=\mbox{constant}$ for all $i\ne j$?
If so, then the operators
$E_i=\frac{1}{d}|\psi_i\rangle\langle\psi_i|$ can be shown to
form the elements of a tomographically complete positive
operator-valued measure, often dubbed a SIC-POVM. Interestingly,
despite the elementary feel to this question---i.e., it {\it seems\/}
the sort of thing one might find as an exercise in a linear-algebra
textbook---and the considerable efforts to solve it, the answer
remains elusive. One should ask: why is there such interest in the
question in the first place? Unfortunately it is hard to deny the
impression that much of the motivation stems from little more than
the mathematical challenge---not really a good physical reason for
months of effort. In this paper, we address the issue of whether
there are other, more physical motivations for seeking an answer to the existence of such states.

Indeed, there are already notable motivations from applied
physics---e.g., uses of these states in quantum cryptography
\cite{apps}. Here, however, we go further by deriving a few
previously unobserved properties that single out SIC-sets as
particularly relevant to the elementary structure of Hilbert space.
These properties make no direct use of the original defining symmetry
for SICs, and instead imply them.  It is our hope that a better
understanding of SIC-sets will lead to new and powerful tool for
studying the structure of quantum mechanics \cite{quantumness}.

Specifically, the paper is as follows. We first demonstrate a geometric sense in
which SIC-sets of projectors $\Pi_i=|\psi_i\rangle\langle\psi_i|$, if they
exist, are as close as possible to being an orthonormal basis on the cone of
nonnegative operators. This complements the frame theoretic version of the same
question proved by Scott \cite{Scott}. The two results together clinch the idea
that SIC-sets are uniquely singled out as a stand-in for orthonormal bases on
the space of density operators, and we take the opportunity to express the
structure of pure states with respect to these preferred bases. Thereafter we
focus on the case of Weyl-Heisenberg (WH) covariant SIC-sets in prime
dimensions. In prime dimensions, complete sets of mutually unbiased bases always
exist \cite{Ivanovic} and here we demonstrate a simple expression for them in
terms of the WH unitary operators. We then define a notion of {\it minimum
uncertainty state\/} with respect to the measurement of these bases
\cite{WoottersSussman} and find that the Weyl-Heisenberg SIC-sets, whenever they
exist, consist solely of minimum uncertainty states. This provides a strong
motivation for considering WH SIC-sets as a special finite-dimensional analogue
of coherent states and stresses the further interest of writing the quantum
mechanical state space in terms of such operator bases. Finally, we conclude
with some general remarks and conjecture a significant reduction in the defining
equations for WH SIC-sets.

\section{Quasi-Orthonormal Bases and the Space of Density Operators}
\label{sec:Quasi}

When equipped with the usual Hilbert-Schmidt inner product $\langle
A, B \rangle =\tr (A^ {\dagger} B)$, the set of operators acting on a
$d$-dimensional complex vector space becomes a $d^2$-dimensional
Hilbert space. Suppose we are given $d^2$ operators $A_i$ normalized
so that $\tr(A^2_i) = 1$.  Under what conditions can the $A_i$ be
orthogonal to each other?  For instance, it is known that one can
require the $A_i$ to be Hermitian or unitary and it is not too
restrictive for orthogonality to obtain \cite{Schwinger}.  However,
requiring the operators to be positive semi-definite is another
matter.

We would like to measure ``how orthonormal'' a set of positive
semi-definite $A_i$ can be. A natural class of measures for this is
\begin{equation}
K_t= \sum_{i \neq j} |\langle A, B \rangle|^t = \sum_{i \neq j}
\bigl(\tr(A_i A_j)\bigr)^t\;, \label{eq:KDef}
\end{equation}
for any real number $t\ge1$. Clearly, one will have an orthonormal basis if and
only if $K_t=0$ for this sum of $N=d^4-d^2$ terms. But for positive
semi-definite $A_i$, as we will prove, it turns out that
\begin{equation}
K_t \ge \frac{d^2(d-1)}{(d+1)^{t-1}}\;. \label{eq:KBound}
\end{equation}
In other words, it is impossible to choose an orthonormal basis all
of whose elements lie in the positive cone of operators. The natural
question to ask is, is there a set of $d^2$ $A_i$ for which $K_t$
achieves the lower bound for $t>1$ (for $t=1$, any set of $A_i/d$
that forms a POVM would saturate the bound for $K_1$)? If so, we will refer to such a set as a
quasi-orthonormal basis.

Note that the linear independence of the $A_i$ does not have to be
imposed as a separate requirement. An orthonormal set of vectors is
automatically linearly independent. Similarly with quasi-orthonormal
sets: requiring $K_t$ to achieve its lower bound forces linear
independence. This will follow from one condition for achieving
equality---that $\langle A_i, A_j \rangle=\mbox{constant}$ when
$i\ne j$ \cite{Caves}.

We will first prove (\ref{eq:KBound}) for the case $t=1$, using a
special instance of the Cauchy-Schwarz inequality: namely, if
$\lambda_r$ is any set of $n$ real numbers, $\sum_r \lambda_r^2 \ge
\frac{1} {n} \left(\sum_r \lambda_r\right)^2$, with equality if and
only if $\lambda_1 = \dots = \lambda_n$. Let $G$ be a positive
semi-definite operator defined by $G = \sum_i A_i$, and note that
$\tr A_i \ge 1$ because $\tr A_i^2 = 1$. Applying the Schwarz
inequality to the eigenvalues of $G$, we find $\tr G^2 \ge
\frac{1}{d}\big(\tr G\big)^2 \ge d^3$, or equivalently, $K_1 \ge
d^3-d^2$. Equality obtains if and only if $G = d I$ and $\tr A_i=1$
for all $i$, i.e. $A_i$ are all rank-1 projectors.

Next, let $f(x) = x^t$, then $f(x)$ is a strictly convex function
for $t>1$. By writing $K_t=\sum_{i \neq j} f\bigl(\tr(A_i
A_j)\bigr)$, and applying Jensen inequality, we complete the proof
for (\ref{eq:KBound}). In Jensen inequality, the equality holds if
and only if $\tr(A_i A_j)$ is a constant for all $i \neq j$.

Putting all this together, we conclude that the necessary and
sufficient conditions for the $A_i$ to constitute a
quasi-orthonormal basis are:
\begin{enumerate}
\item Each $A_i$ is a rank-1 projector.
\item $\tr(A_i A_j) = \frac{1}{d+1}$ for all $i \neq j$.
\end{enumerate}
These are the same as the conditions for the operators $A_i$ to
constitute a SIC-set.  In particular they imply $\sum_i A_i = d I$
and, via \cite{Caves}, that the $A_i$ must be linearly independent.

As an aside, let us point out that this derivation sheds light on
the meaning of the ``second frame potential'' $\Phi=\sum_{i, j}
\bigl| \langle \psi_i |\psi_j\rangle\bigr|^4$ introduced by Renes
{\it et al.}~\cite{Renes} to aid in finding SIC-POVMs numerically.
Beforehand its motivation seemed to be only the heuristic of
constrained particles distancing themselves from each other because
of an interaction potential---an idea that has its roots in
Ref.~\cite{Benedetto}. Now, one sees that whenever
$A_i=|\psi_i\rangle\langle\psi_i|$, $\Phi = K_2 + d^2$---i.e.,
$\Phi$ is essentially our own orthonormality measure in this case.
(\ref{eq:KBound}) implies that $\Phi \ge 2 d^3/(d+1)$, which was
proved by different means in Ref.~\cite{Renes}.

Returning to the general development, what we have shown is that
SIC-sets $\Pi_i=|\psi_i\rangle\langle\psi_i|$, $i=1,\ldots,d^2$, play
a special role in the geometry of the cone of positive operators, and
by implication, the convex set of quantum states in general---i.e.,
the density operators.  For the purpose of foundational
studies---particularly ones of the quantum Bayesian variety
\cite{quantumness,BayesianLump}---it is worthwhile recording what
this convex set looks like from this perspective.

Imagine introducing $E_i=\frac{1}{d}\Pi_i$ as a canonically given
quantum measurement.  For a quantum state $\rho$, the outcomes will
occur with probabilities $p(i)=\tr\rho E_i$.  Using the fact that
$\rho$ has a unique expansion in terms of the $\Pi_i$, one can work
out that
\begin{equation}
\rho=\sum_i \left((d+1)p(i)-\frac{1}{d}\right)\Pi_i\;.
\label{mannequin}
\end{equation}
On the other hand, one might imagine taking the vector of
probabilities $p(i)$ as the more basic specification, and the density
operator $\rho$ as a convenient, but derivative, specification of the
quantum state.  Requisite to doing this, one must have an
understanding of the {\it allowed\/} probabilities $p(i)$---not every
choice of probabilities in Eq.~(\ref{mannequin}) will give rise to
positive semi-definite $\rho$. This can be done conveniently by
taking note of the characterization of pure quantum states
demonstrated in Ref.~\cite{JonesLinden}.  Remarkably, it is enough to
specify that any Hermitian operator $M$ satisfy only two trace
conditions, $\tr M^2=1$ and $\tr M^3=1$, to insure that it be a
rank-1 projection operator---i.e., a pure state.  In terms of the
$p(i)$ these two conditions become:
\begin{eqnarray}
\sum_i p(i)^2 &=& \frac{2}{d(d+1)}
\label{Jack}
\\
\sum_{i,j,k} c_{ijk}\, p(i)p(j)p(k) &=& \frac{d+7}{(d+1)^3}\;,
\label{Jill}
\end{eqnarray}
where $c_{ijk} = \mbox{Re}\, \tr\!\big( \Pi_i\Pi_j\Pi_k\big)$. The
full set of quantum states is thus the convex hull of the solutions
to Eqs.~(\ref{Jack}) and (\ref{Jill}). One nice thing about
Eq.~(\ref{Jill}) is that it makes transparent the algebraic structure
lying behind these special probability vectors:  For, one can use the
coefficients $c_{ijk}$ as structure coefficients in {\it defining\/}
the anticommutator on the space of operators.

\section{Weyl-Heisenberg SIC-Sets and Minimum Uncertainty States}
\label{sec:WHquasiONBs}
A particularly important class of SIC-sets are those covariant under
the action of the Weyl-Heisenberg (WH) group. Most of the SIC-sets
constructed to date are of this variety \cite{Zauner,Renes,JMP,SICGeneral,Minsk},
and these structures have many pleasing properties. In this Section,
we derive a few new properties of such sets and show a sense in which
their elements can be called {\it minimum uncertainty states\/} when
$d$ is prime.

Choose a standard basis $|0\rangle, \dots, |d-1\rangle$ in ${\cal
H}_d$. Let $ Z |j\rangle  = \omega^{j} | j \rangle $ and $ X |
j\rangle = | j +1 \rangle$, where $\omega=e^{2 \pi i /d}$ and the
addition in $| j+1\rangle $ is modulo $d$. The WH displacement
operators are defined by $D_{\mathbf{r}} = \tau^{r_1 r_2} X^{r_1}
Z^{r_2}$ for $\mathbf{r} = (r_1, r_2)$ and $\tau= - e^{i \pi/d}$. A
WH covariant SIC-set is constructed by taking a single normalized
vector $|\psi\rangle$ and acting on it with the unitaries
$D_{\mathbf{r}}$. As $r_1, r_2$ range over the integers $0, \dots,
d-1$ this gives $d^2$ vectors $|\psi_{\mathbf {r}}\rangle=
D_{\mathbf{r}} | \psi\rangle$.  If a $|\psi\rangle$ exists such that
\begin{equation}
\langle \psi | D_{\mathbf{r}}| \psi \rangle  =
\begin{cases} 1 \qquad & \mathbf{r} = \bm{0} \\
\frac{e^{i \theta_{\mathbf{r}} } }{\sqrt{d+1}} \qquad & \mathbf{r}
\neq \bm{0}\;,
\end{cases}
\label{eq:WHGram}
\end{equation}
the $|\psi_{\mathbf{r}}\rangle \langle \psi_{\mathbf{r}}|$ form a
SIC-set and we call $|\psi\rangle$ fiducial.

We can write Eq.~(\ref{eq:WHGram}) in a more convenient form by
working things out directly in terms of components.  We have $\langle
\psi | D_\mathbf{r} | \psi \rangle = \tau^{r_1 r_2} \sum_j \omega^{j
r_2} \psi^{*}_{j+r_1} \psi^{\vphantom{*}}_j$ where $\psi_j = \langle
j | \psi\rangle$. Consequently, taking a Fourier transform,
\begin{equation}
\frac{1}{d} \sum_{r_2} \omega^{k r_2} \left| \langle \psi |
D_{\mathbf {r}} |\psi\rangle \right|^2 = \sum_{j}
\psi^{\vphantom{*}}_j \psi^{*}_{j+k} \psi^{*}_{j+r_1}
\psi^{\vphantom{*}}_{j+k+r_1}. \label{eq:FTA}
\end{equation}
Replacing $\left|\langle \psi | D_{\mathbf{r}} | \psi \rangle \right|
^2$ with $(d \delta_{\mathbf{r}, \bm{0}}+1)/(d+1)$  we deduce that $|
\psi\rangle$ is a fiducial vector if and only if
\begin{equation}
\sum_{j} \psi^{\vphantom{*}}_j \psi^{*}_{j+k} \psi^{*}_{j +l}
\psi^{\vphantom{*}}_{j+k+l}  = \frac{1}{d+1} \left( \delta_{k 0} +
\delta_{l 0}\right)\;. \label{eq:NewWHConds}
\end{equation}
In the equations $\left|\langle \psi | D_{\mathbf{r}} | \psi \rangle
\right|^2 = (d \delta_{\mathbf{r}, \bm{0}}+1)/(d+1)$, the left hand
sides are sums of terms quartic in the components, multiplied by
powers of $\omega$.  In Eqs.~(\ref{eq:NewWHConds}) the powers of $
\omega$ have disappeared; this makes them easier to work with.
Similarly, note that the frame potential $\Phi$ can be written in an
$\omega$-free form.  Starting from Eq.~(\ref{eq:FTA}), it follows that
\begin{equation}
%$
\Phi= d^3 \sum_{k,l}\Bigl| \sum_j \psi^{\vphantom{*}}_j
\psi^{*}_{j+k} \psi^{*}_{j +l} \psi^{\vphantom{*}}_{j+k+l}
\Bigr|^2\;.
%$
\end{equation}
Finally, let us note the important case of
Eqs.~(\ref{eq:NewWHConds}) where $l=0$. The only terms appearing are
then, in terms of probabilities $p_j = |\psi_j|^2:$
\begin{equation}
\sum_j p_j p_{j+k} = \frac{1}{d+1}(1+\delta_{k 0})\;.
\label{eq:GpIA}
\end{equation}

%%%%%%%%%%%%%%%%%%%%%%%%%%%%%%%%%%%%%% (Start of old ``Minimum Uncertainty State'' section.)

Let us now specialize to prime dimension and work toward expressing
the fiduciality condition Eq.~(\ref{eq:NewWHConds}) in terms of
probabilities of measurement outcomes for a complete set of $d+1$
mutually unbiased bases (MUBs). Assume $d>2$ (the $d=2$ case requires
special treatment, though our conclusions continue to hold).

Consider the matrix
$ F=\bigl(\begin{smallmatrix} { \alpha} & \beta \\
\gamma & \delta\end{smallmatrix}\bigr) $ where $\alpha, \beta,
\gamma, \delta$ are nonnegative integers $< d$ such that $\det F = 1
\; \text{mod} \; d$. We will refer to such matrices as symplectic
matrices.  Let $U$ be the unitary
\begin{equation}
U =
\begin{cases} \frac{1}{\sqrt{d}} \sum_{j,k} \tau^{\beta^{-1} (\alpha
k^2 - 2 j k + \delta j^2)} |j\rangle\langle k | \quad & \beta \neq 0 \\
\sum_{j}\tau^{\alpha \gamma j^2} |\alpha j\rangle \langle j | \quad &
\beta = 0\;.
\end{cases}
\label{eq:RDefA}
\end{equation}
Then it can be shown \cite{JMP} that
\begin{equation}
U D_{\mathbf{r}} U^{\dagger} = D_{F \mathbf{r}}
\label{eq:RDefB}
\end{equation}
for all $\mathbf{r}$. Moreover, if $U'$ is any other unitary with
this property, then $U' = e^{i \phi} U$ for some phase $e^{i \phi}$.
In these expressions all arithmetical operations on the indices are
mod $d$.  In particular $\beta^{-1}$ is the unique positive integer
less than $d$ such that $\beta^{-1}
\beta=1 \; \text{mod}\; d$.

If $\beta\neq 0$, it can be seen that $|\langle j | U | k\rangle | =
d^ {-1/2}$ for all $j, k$.  So we can use symplectic transformations
to generate MUBs.  For definiteness take $V$, $W$ to be the unitaries
corresponding to $\bigl(\begin{smallmatrix} 1 & 1 \\ 0 &
1\end {smallmatrix}\bigr)$ and $\bigl(\begin{smallmatrix} 0 & 1 \\
-1
& 0\end{smallmatrix}\bigr)$ and define
\begin{equation}
| m , j\rangle =
\begin{cases} V^m | j\rangle \qquad & m = 0, 1, \dots , d-1 \\
W | j\rangle \qquad & m = \infty\;.
\end{cases}
\label{eq:MUBDef}
\end{equation}
This gives us a full set of $d+1$ MUBs labeled by  $m$.

These bases are essentially the only ones obtainable from the
standard basis by symplectic transformations.   To see this let $F$
be an arbitrary symplectic matrix with  $\beta \neq 0$, and let $U$
be the corresponding unitary.     We have
\begin{equation}
F = \begin{pmatrix} \alpha & \beta \\ \gamma & \delta\end{pmatrix}=
\begin{cases}
\begin{pmatrix} 1 & 1 \\ 0 & 1\end{pmatrix}^{\beta \delta^{-1}}
\!\!\begin{pmatrix} \delta^{-1} & 0 \\ \gamma & \delta \end{pmatrix}
\qquad & \delta \neq   0
\\
\begin{pmatrix} 0 & 1 \\ -1 & 0\end{pmatrix} \begin{pmatrix} \beta^
{-1} & 0 \\ \alpha & \beta \end{pmatrix} \qquad & \delta = 0\;.
\end{cases}
\end{equation}
But, unitaries corresponding to matrices with a $0$ in the top
right-hand position merely permute and re-phase the elements of the
standard basis. So the basis $U|0\rangle, \dots, U| d-1\rangle$
coincides with one of the bases $|m, 0\rangle , \dots | m,
(d-1)\rangle$ up to permutation and re-phasing. We have $| j \rangle
\langle j | = (1/d) \sum_r\omega^{- j r} D_ {(0,r)}$.  In view of
Eqs.~(\ref{eq:RDefB}) and~(\ref{eq:MUBDef}) this means
\begin{equation}
| m , j\rangle \langle m, j | =
\begin{cases}
\frac{1}{d} \sum_r \omega^{-j r} D_{(m r, r)} \quad & m\neq \infty \\
\frac{1}{d} \sum_r \omega^{-j r} D_{(r,0)} \quad & m = \infty\;.
\end{cases}
\end{equation}
Now consider an arbitrary $|\psi\rangle$, and let $p_{m, j}=| \langle
m, j | \psi\rangle|^2$. It follows from the above that
\begin{equation}
\sum_{j} p_{m, j} p_{m, j+ k} =
\begin{cases}
\frac{1}{d} \sum_{r} \omega^{k r} \left| \langle \psi | D_{(m r, r)}
| \psi \rangle \right|^2  & m\neq \infty
\\
\frac{1}{d} \sum_{r} \omega^{k r} \left| \langle \psi | D_{(r, 0)} |
\psi \rangle \right|^2  & m= \infty\;,
\end{cases}
\end{equation}
from which one sees that $|\psi\rangle$ is fiducial if and only if
\begin{equation}
\sum_{j} p_{m, j} p_{m,j+k}= \frac{1}{d+1}(1+\delta_{k 0})\;.
\label{eq:GpIB}
\end{equation}
Thus, instead of imposing all of the Eqs.~(\ref{eq:NewWHConds}) on
the components in the standard basis, we can impose just the special
case Eqs.~(\ref{eq:GpIA}) for each MUB separately.

We are now ready to derive the minimum uncertainty property.  Consider an arbitrary state $|\psi\rangle$ and a
measurement of one of the bases $m$ in a complete set of MUBs.  This
will give rise to outcomes with a probability distribution $p_{m,j}$.
We will quantify the uncertainty in the outcomes by the quadratic
R\'{e}nyi entropy~\cite{WoottersSussman,AczelDaroczy}
\begin{equation}
H_m = - \log_2 \Bigl(\sum_{j} p^{2}_{m,j}\Bigr)\;.
\end{equation}
This measure and the measure related to it by deleting the logarithm
seem to be playing a widening role in quantum information studies
\cite{BruknerZeilinger}. In particular, it is one of the most
important measures for quantifying an eavesdropper's information in
quantum cryptography \cite{SomethingFromNorbert}.

A minimum uncertainty state is one that minimizes the total
uncertainty, $T=\sum_{m} H_m$. To see the conditions for this, we
first appeal to the fact that \cite{BruknerZeilinger,Larsen}
\begin{equation}
\sum_{m, j} p^2_{m, j} = 2
\label{eq:SumpjmSquared}
\end{equation}
for any $|\psi\rangle$. From the convexity of the logarithm, it
follows that for any sequence of positive numbers $\lambda_1, \dots,
\lambda_n$, $(1/n)\sum_{j} \log_2 \lambda_j \le  \log_2
\left((1/n)\sum_j \lambda_j\right)$ with equality if and only if
$\lambda_1 = \dots = \lambda_n$.  Thus, in view of
Eq.~(\ref{eq:SumpjmSquared})
\begin{equation}
T \ge (d+1) \log_2\! \left(\frac{d+1}{2}\right)\;,
\end{equation}
with equality if and only if $\sum_j p^2_{m,j} = 2/(d+1)$ for all
$m$.  Comparing this with Eq.~(\ref{eq:GpIB}), one sees that every WH
fiducial vector achieves the lower bound and is therefore a minimum
uncertainty state.  Unfortunately, the theorem does not go the other
way:  it is not always the case that every minimum uncertainty state
is a fiducial state.

\section{Discussion}

We have argued that the SIC-sets in general dimensions, and more
specifically the Weyl-Heisenberg covariant ones (at least in prime
dimensions), are particularly interesting structures in Hilbert space
geometry.  Thus they call out for a better understanding, and we hope
this paper is an advertisement for that cause.

There are clearly many more things that can be asked.  For instance, are SIC fiducial states also minimum uncertainty states in
prime-power dimensions?  In non-prime-power dimensions, we face the
problem that complete sets of MUBs may not even exist.
In that case,
can one generalize the definition of a MUB in such a
way that SIC fiducial states continue to be minimum uncertainty
states? With regard to the property of quasi-orthonormality, is it
possible to find {\it non}-Weyl-Heisenberg SIC-sets for which
Eq.~(\ref{Jill}) takes a nicer form?
Further, with regard to the property of quasi-orthonormality, is it
possible to find {\it non}-Weyl-Heisenberg SIC-sets for which
Eq.~(\ref{Jill}) takes a nicer form?  For instance, if the $\Pi_i$
were actually orthogonal, then one would have
$\tr\big(\Pi_i\Pi_j\Pi_k\big)=\delta_{ij}\delta_{jk}\delta_{ki}$. How
close can one come to a form like this with a SIC-set?  Many
questions like this loom.

But most looming is the question of whether SIC-sets exist in
arbitrary dimension.  In this regard, let us close with a
conjecture. The necessary and sufficient conditions for
$|\psi\rangle$ to be a Weyl-Heisenberg fiducial vector are captured
in Eq.~(\ref{eq:NewWHConds}), which  represents $d^2$ simultaneous
equations.  However, we have noted numerically (up to $d=28$) that
satisfying roughly $\frac{3}{2}d$ of these equations is enough to
imply the rest.  For instance, it appears to be necessary to satisfy
Eq.~(\ref{eq:NewWHConds}) only for the cases $k=0,1,2$ with
$l=0,\ldots \ulcorner d/2\urcorner$.  Perhaps this always holds
true.

This research was supported in part by the U.~S. Office of Naval Research (Grant No.\ N00014-09-1-0247).


\begin{thebibliography}{99}
\bibitem{Zauner}
 G.~Zauner, Ph.D.
Thesis, University of Vienna (1999).

\bibitem{Caves}
C.~M. Caves, (1999); see {\tt
http://info.phys.unm.edu/$\sim$ caves/reports/infopovm.pdf}.

\bibitem{Renes}
J. M.~Renes {\it et al.}, J.\ Math.\ Phys.\  \textbf{45}, 2171 (2004).

\bibitem{JMP}
D. M.~Appleby, J.\ Math.\ Phys.\ \textbf{46}, 052107 (2005).

\bibitem{SICGeneral}
M.~Grassl, quant-ph/0406175; W.~K. Wootters, Found.\ Phys.\
\textbf{36}, 112 (2006); A.~Klappenecker {\it et al.}, quant-ph/
0503239; S.~T. Flammia, J. Phys.\ A \textbf{39} 13483 (2006).

\bibitem{ScottA}
A. J.~Scott and M.~Grassl, quant-ph/0910.5784

\bibitem{Minsk}
D. M.~Appleby, quant-ph/0611260.

\bibitem{apps}
C.~A. Fuchs and M.~Sasaki, Quant.\ Inf.\ Comp.\ \textbf{3}, 277
(2003); J.~\v{R}eh\'a\v{c}ek {\it et al.}, Phys.\ Rev. A \textbf{70},
052321 (2004); J.~M. Renes, Quant.\ Inf.\ Comp.\ \textbf{5}, 80
(2005); I.~H. Kim, quant-ph/0608024.

\bibitem{quantumness}
C.~A. Fuchs, Quant.\ Inf.\ Comp.\ \textbf{4}, 467 (2004); C.~A.
Fuchs, quant-ph/0205039; C.~A. Fuchs and R.~Schack,
quant-ph/0906.2187.

\bibitem{Scott}
A.~J. Scott, J. Phys.\ A \textbf{39}, 13507 (2006).

\bibitem{Ivanovic}
I.~D. Ivanovic, J. Phys.\ A \textbf{14}, 3241 (1981).

\bibitem{WoottersSussman}
W.~K. Wootters \& D.~M. Sussman, quant-ph/0704.1277.

\bibitem{Schwinger}
J.~Schwinger, Proc.\ Nat.\ Acad.\ Sci.\ \textbf{46}, 570 (1960).

\bibitem{Benedetto}
J.~J. Benedetto and M. Fickus, Adv.\ Comp.\ Math.\ \textbf{18}, 357
(2003).

\bibitem{BayesianLump}
C.~M. Caves, C.~A. Fuchs and R. Schack, quant-ph/ 0608190; C.~A.
Fuchs, and R. Schack, quant-ph/0404156; D.~M. Appleby, Found.\ Phys.\
\textbf{35}, 627 (2005).

\bibitem{JonesLinden}
N.~S. Jones and N. Linden, Phys.\ Rev.\ A \textbf{71}, 012324 (2005); S.~T. Flammia, unpublished (2004).

\bibitem{BruknerZeilinger}
C.~Brukner and A.~Zeilinger, Phys.\ Rev.\ A \textbf{63}, 022113
(2001).

\bibitem{Larsen}
U. Larsen, J.\ Phys.\ A {\bf 23}, 1041 (1990).

\bibitem{AczelDaroczy}
J.~Acz\'el and Z.~Dar\'oczy, {\sl On Measures of Information and
Their Characterizations}, (Academic, New York, 1975).

\bibitem{SomethingFromNorbert}
N.~L\"utkenhaus, Phys.\ Rev.\ A, \textbf{59}, 3301 (1999).

\end{thebibliography}
\end{document}